\documentclass[aip,jcp,preprint,amsmath]{revtex4-1}
\usepackage{graphicx,dcolumn,bm,color,microtype,multirow,mathtools,setspace,hyperref}
\usepackage[version=3]{mhchem}

\newcommand{\Ec}{E_\text{c}}
\newcommand{\EHF}{E_\text{HF}}
\newcommand{\Eex}{E_\text{exact}}

\newcommand{\EP}{E_c^\text{MP2}}
\newcommand{\EPP}{E_c^\text{MP3}}

\newcommand{\Esoft}{E_c^\text{soft}}

\newcommand{\G}{\Gamma}
\newcommand{\mL}{\mathcal{L}}
\newcommand{\mE}{\mathcal{E}}
\newcommand{\mO}{\mathcal{O}}
\newcommand{\mR}{\mathcal{R}}

\newcommand{\mc}{\multicolumn}

\newcommand{\Eh}{E_\text{h}}
\newcommand{\mEh}{\text{m}E_\text{h}}

\newcommand{\Req}{R_\text{eq}}
\newcommand{\Rts}{R_\text{ts}}
\newcommand{\dip}{$\langle x \rangle$}
\newcommand{\rms}{$\sqrt{\langle x^2 \rangle}$}

\begin{document}

\title{Chemistry in One Dimension}

\author{Pierre-Fran{\c c}ois Loos}
\email{Corresponding author: pf.loos@anu.edu.au}
\author{Caleb J. Ball}
\author{Peter M. W. Gill}
\email{peter.gill@anu.edu.au}
\affiliation{Research School of Chemistry, Australian National University, Canberra ACT 0200, Australia}

\begin{abstract}
We report benchmark results for one-dimensional (1D) atomic and molecular systems interacting via the Coulomb operator $|x|^{-1}$. 
Using various wavefunction-type approaches, such as Hartree-Fock theory, second- and third-order M{\o}ller-Plesset perturbation theory and explicitly correlated calculations, we study the ground state of atoms with up to ten electrons as well as small diatomic and triatomic molecules containing up to two electrons.
A detailed analysis of the 1D helium-like ions is given and the expression of the high-density correlation energy is reported.
We report the total energies, ionization energies, electron affinities and other interesting properties of the many-electron 1D atoms and, based on these results, we construct the 1D analog of Mendeleev's periodic table. 
We find that the 1D periodic table contains only two groups: the alkali metals and the noble gases. 
We also calculate the dissociation curves of various 1D diatomics and study the chemical bond in \ce{H2+}, \ce{HeH^2+}, \ce{He2^3+}, \ce{H2}, \ce{HeH+} and \ce{He2^{2+}}. 
We find that, unlike their 3D counterparts, 1D molecules are primarily bound by one-electron bonds.
Finally, we study the chemistry of \ce{H3+} and we discuss the stability of the 1D polymer resulting from an infinite chain of hydrogen atoms.
\end{abstract}

	
\maketitle

\section{1D chemistry}
Chemistry in one dimension (1D) is interesting for many experimental and theoretical reasons, but also in its own right.
Experimentally, 1D systems can be realized in carbon nanotubes, \cite{SaitoBook, Egger98, Bockrath99, Ishii03, Shiraishi03} organic conductors, \cite{Schwartz98, Vescoli00, Lorenz02, Dressel05, Ito05} transition metal oxides, \cite{Hu02} edge states in quantum Hall liquids, \cite{Milliken96, Mandal01, Chang03} semiconductor heterostructures, \cite{Goni91, Auslaender00, ZaitsevZotov00, Liu05, Steinberg06} confined atomic gases, \cite{Monien98, Recati03, Moritz05} and atomic or semiconducting nanowires. 
Theoretically, Burke and coworkers \cite{Wagner12, Stoudenmire12} have shown that 1D systems can be used as a ``theoretical laboratory'' to study strong correlation in ``real'' three-dimensional (3D) chemical systems within density-functional theory. \cite{ParrBook} Herschbach and coworkers calculated the ground-state electronic energy of 3D systems by interpolating between exact solutions for the limiting cases of 1D and infinite-dimensional systems. \cite{Doren85, Loeser86a, Loeser86b} 

However, all these authors eschewed the Coulomb operator $1/|x|$ because of its strong divergence at $x=0$.  
For example, Burke and coworkers \cite{Wagner12, Stoudenmire12} used a softened version of the Coulomb operator $1/\sqrt{x^2+1}$ to study 1D chemical systems, such as light atoms (\ce{H}, \ce{He}, \ce{Li}, \ce{Be}, \ldots), ions (\ce{H-}, \ce{Li+}, \ce{Be+}, \ldots), and diatomics (\ce{H2+} and \ce{H2}).
Herschbach and coworkers have worked intensively on the 1D He atom,\cite{Rosenthal71, Foldy76, Foldy77, Nogami76} replacing the usual Coulomb inter-particle interactions with the Dirac delta function $\delta(x)$. \cite{Loeser85, Herschbach86, Doren87, Loeser87b} 
There are few studies using the true Coulomb operator and most of these focus on non-atomic and non-molecular systems. \cite{Astrakharchik11, Lee11a, QR12, UEG1D, Ringium13, GLDA1, GLDA1w}
We prefer the Coulomb operator because, although it is not the solution of the 1D Poisson equation, it pertains to particles that are \emph{strictly} restricted to move in a one-dimensional sub-space of three-dimensional space.

The first 1D chemical system to be studied was the H atom by Loudon. \cite{Loudon59} 
Despite its simplicity, this model has been useful for studying the behavior of many physical systems, such as Rydberg atoms in external fields \cite{Burnett93, Mayle07} or the dynamics of surface-state electrons in liquid helium \cite{Nieto00, Patil01} and its potential application to quantum computing. \cite{Platzman99, Dykman03} 
Most work since Loudon has focused on one-electron ions\cite{Loudon59, Andrews76, NunezYepez87b, Boya88, Mineev04, deOliveira09, Abramovici09, NunezYepez11}  and, to the best of our knowledge, no calculation has been reported for larger chemical systems.
In part, this can be attributed to the ongoing controversy concerning the mathematical structure of the eigenfunctions. \cite{Imbo85, NunezYepez87a, NunezYepez89, Moshinsky93, Xianxi97, Connolly07, NunezYepez11} 
Although debates about the parities and boundedness of the eigenfunctions continue, we will assume in the present study  that nuclei are impenetrable. \cite{Moshinsky93, Newton94, NunezYepez11}

In Sec.~\ref{sec:atoms} and Sec.~\ref{sec:molecules}, we report electronic structure calculations for 1D atomic and molecular systems using the Coulomb operator $1/|x|$.  Sec.~\ref{sec:molecules} discusses several diatomic systems, the chemistry of \ce{H3+} and an infinite chain of 1D hydrogen atoms.

Because of the singularity of the Coulomb interaction in 1D, the electronic wave function has nodes at all points where two electrons touch. \cite{Mitas06}
As a result, all 1D systems are spin-blind\cite{Astrakharchik11, Lee11a, QR12, Ringium13} and we are free to assume that all electrons have the same spin.
This also means that the Pauli exclusion principle forbids \emph{any} two electrons in 1D to have the same quantum state and, in independent-electron models such as Hartree-Fock (HF) theory,\cite{SzaboBook} orbitals have a maximum occupancy of one. 
Unless otherwise stated, atomic units are used throughout:  total energies in hartrees ($\Eh$), correlation energies in millihartrees ($\mEh$) and bond lengths in bohrs.

\section{Theory}

\subsection{Notation}
Because of the impenetrability of the 1D Coulomb potential, \cite{Newton94, NunezYepez11} electrons cannot pass from one side of a nucleus to the other and are trapped on domains which are either rays (to the left or right of the molecule) or line segments (between nuclei).
The resulting domain occupations lead to families of states which can be defined by specifying the occupied orbitals in each domain. 
For example, the notation \ce{_$i$A$_j^{Z_\text{A}-2}$} denotes an atom \ce{A} of nuclear charge $Z_\text{A}$ whose $i$th left orbital and $j$th right orbitals are (singly) occupied.  
Likewise, \ce{A$_{i,j}^{Z_\text{A}-2}$} indicates an atom with two electrons, in the $i$th and $j$th orbitals, on the right side of the nucleus. 
1D molecules can be similarly described.  For example, \ce{A$_{1,4}$B$_1^{Z_\text{A}+Z_\text{B}-3}$} denotes a diatomic in which two electrons are between the nuclei and one electron is on the \ce{B} side of the molecule. 
When consecutive orbitals are occupied in the same domain, we use dashes.  For example, \ce{A$_{1\text{--}3}$B$^{Z_\text{A}+Z_\text{B}-3}$} implies that the three lowest orbitals between the nuclei are occupied.

\subsection{Computational details}
We have followed the methods developed by Hylleraas \cite{Hylleraas29, Hylleraas64} and James and Coolidge \cite{James33} to compute the exact or near-exact energies $\Eex$ of one-, two-, and three-electron systems.  We have written a standalone program\cite{HF1D} called \textsc{Chem1D} to perform HF and M{\o}ller-Plesset perturbation theory calculations\cite{SzaboBook} on arbitrary 1D atomic and molecular systems.

\begin{figure}[htbp]
	\includegraphics[width=0.9\textwidth]{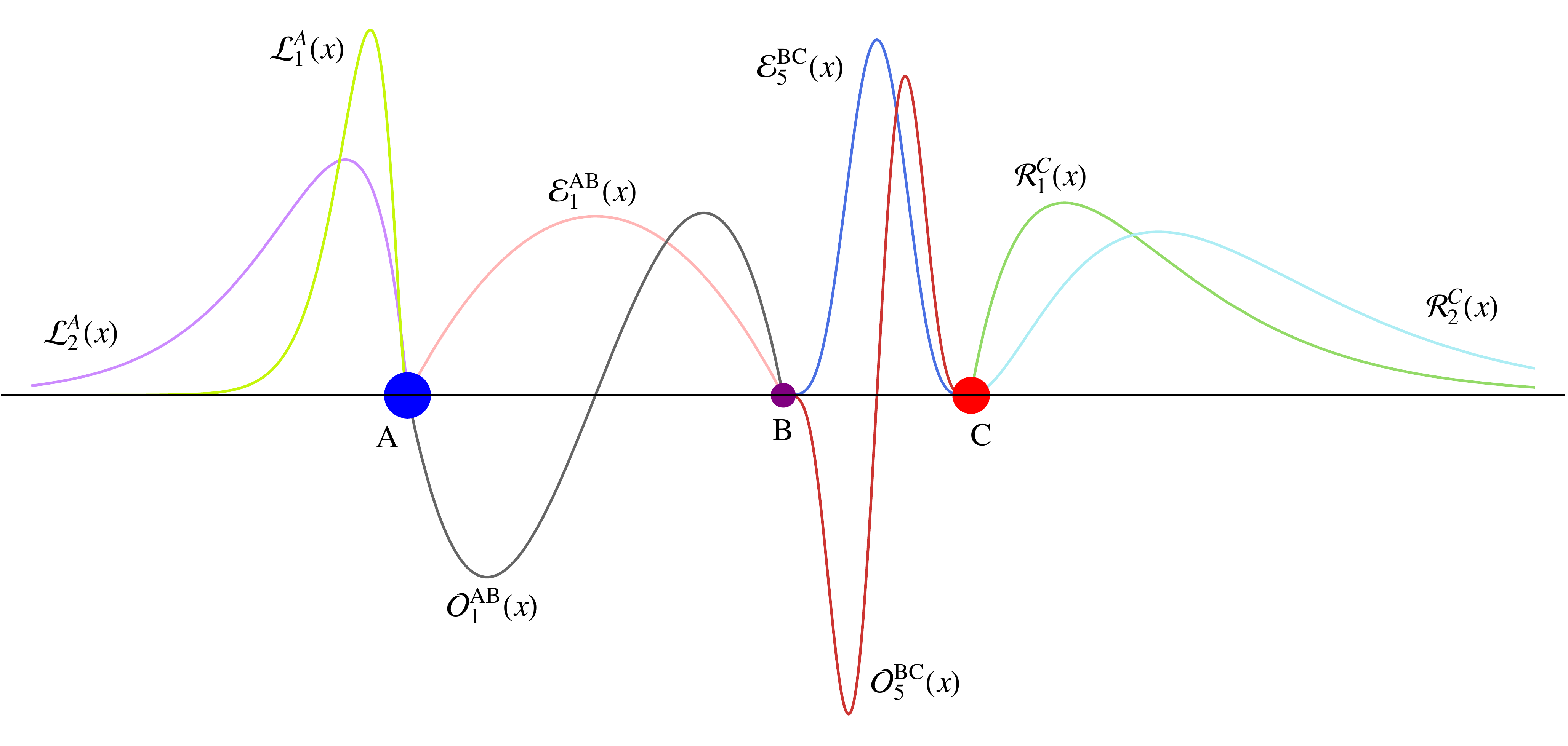}
	\caption{
	\label{fig:HFbasis}
	Orbital basis functions in a triatomic molecule \ce{ABC}}
\end{figure}

All our atomic and molecular calculations use a normalized basis of exponentials on the ray to the left of the leftmost nucleus,
\begin{equation}
	\mL_k^A(x) = 2 k^3 \alpha^{3/2} (A-x) e^{-k^2\alpha(A-x)},
\end{equation}
exponentials on the ray to the right of the rightmost nucleus,
\begin{equation}
	\mR_k^B(x) = 2 k^3 \alpha^{3/2} (x-B) e^{-k^2\alpha(x-B)},
\end{equation}
even polynomials on the line segment between adjacent nuclei,
\begin{equation}
\label{E}
	\mE_k^{AB}(x) = \sqrt{\frac{2/\pi^{1/2}}{R_{AB}} \frac{\G(2k+3/2)}{\G(2k+1)}}\ (1-z^2)^k,
\end{equation}
and odd polynomials on the line segment between adjacent nuclei,
\begin{equation}
	\mO_k^{AB}(x) = \sqrt{\frac{4/\pi^{1/2}}{R_{AB}} \frac{\G(2k+5/2)}{\G(2k+1)}}\ z(1-z^2)^k,
\end{equation}
where $z = (A+B-2x)/(A-B)$, $R_{AB} = |A-B|$ and $\G$ is the Gamma function. \cite{NISTbook}
By including only positive integer $k$, we ensure that the orbitals vanish at the nuclei.
Some of these basis functions are shown in Fig.~\ref{fig:HFbasis}.
Full details of these calculations will be reported elsewhere. \cite{HF1D}

The HF eigenvalues and orbitals yield\cite{SzaboBook} the second- and third-order M{\o}ller-Plesset (MP2 and MP3) correlation energies $\EP$ and $\EPP$, and the exact and HF energies yield the correlation energy
\begin{equation}
	\Ec = E_\text{exact} - \EHF.
\end{equation}

\section{
\label{sec:atoms}
Atoms}

\subsection{Hydrogen-like ions}
The electronic Hamiltonian of the 1D H-like ion with nucleus of charge $Z$ at $x = 0$ is
\begin{equation}
	\hat{H} = - \frac{1}{2} \frac{d^2}{dx^2} - \frac{Z}{|x|},
\end{equation}
and this has been studied in great detail. \cite{Loudon59, Andrews76, NunezYepez87b, Boya88, Abramovici09, NunezYepez11}
Its eigenfunctions are
\begin{align}
	\psi_n^+(x)	& = x L_{n-1}^{(1)}(+2Zx/n) \exp(-Zx/n),	\qquad x > 0,	\label{eq:Hpos}		\\
	\psi_n^-(x)	& = x L_{n-1}^{(1)}(-2Zx/n) \exp(+Zx/n),	\qquad x < 0,	\label{eq:Hneg}
\end{align}
where $L_n^{(a)}$ is a Laguerre polynomial \cite{NISTbook} and $n = 1, 2, 3, \ldots$. 
All of these vanish at the nucleus (which is counter-intuitive) and decay exponentially at large $|x|$. 
Curiously, because of nuclear impenetrability, the ground state of the 1D H atom has a dipole moment and \dip\ = $\pm1.5$.

\subsection{
\label{sec:Helike}
Helium-like ions}
The electronic Hamiltonian of the 1D He-like ion is
\begin{equation}
	\Hat{H} = -\frac{1}{2} \left( \frac{\partial^2}{\partial x_1^2}+\frac{\partial^2}{\partial x_2^2} \right) - \frac{Z}{|x_1|} - \frac{Z}{|x_2|} + \frac{1}{|x_1 - x_2|}
\end{equation}
and two families of electronic states can be considered: 
\begin{itemize}
	\item The one-sided \ce{A$_{i,j}^{Z-2}$} family where both electrons are on the same side of the nucleus;
	\item The two-sided \ce{$_i$A$_j^{Z-2}$} family where the electrons are on opposite sides of the nucleus. 
\end{itemize}
Some of the properties of the first ten ions are gathered in Table \ref{tab:Helike}. 

\begin{table*}
\caption{
\label{tab:Helike}
Energies, HOMO-LUMO gaps and radii of the 1D helium-like ions}
\small
\begin{ruledtabular}
\begin{tabular}{ccccccccc}
				&		\mc{2}{c}{Total energy}			&			\mc{4}{c}{Correlation energy}		&	\mc{2}{c}{HF property}		\\
								\cline{2-3}									\cline{4-7}									\cline{8-9}
Ion				&	$-\Eex$			&	$-\EHF$		&	$-\EP$	&  $-\EPP$	&	$-\Ec$	& $-\Esoft$	&	Gap		& 	\rms	 	\\
\hline
\ce{_1H1-}		&	0.646 584		&	0.643 050		&	1.713	&	2.530	&	3.534	&	39		&	0.170		&	2.296		\\
\ce{_1He1}		&	3.245 944		&	3.242 922		&	2.063	&	2.688	&	3.022	&	14		&	1.265		&	0.985		\\
\ce{_1Li1+}		&	7.845 792		&	7.842 889		&	2.235	&	2.733	&	2.903	&	8		&	3.200		&	0.628		\\
\ce{_1Be1^2+}	&	14.445 725		&	14.442 873		&	2.335	&	2.747	&	2.851	&	6		&	5.874		&	0.460		\\
\ce{_1B1^3+}	&	23.045 686		&	23.042 864		&	2.401	&	2.751	&	2.822	&			&	9.294		&	0.364		\\
\ce{_1C1^4+}	&	33.645 661		&	33.642 859		&	2.447	&	2.752	&	2.802	&			&	13.463		&	0.301		\\
\ce{_1N1^5+}	&	46.245 644		&	46.242 855		&	2.481	&	2.751	&	2.789	&			&	18.382		&	0.256		\\
\ce{_1O1^6+}	&	60.845 631		&	60.842 852		&	2.508	&	2.749	&	2.779	&			&	24.050		&	0.223		\\
\ce{_1F1^7+}	&	77.445 621		&	77.442 849		&	2.529	&	2.748	&	2.772	&			&	30.468		&	0.198		\\
\ce{_1Ne1^8+}	&	96.045 613		&	96.042 847		&	2.546	&	2.746	&	2.766	&			&	37.635		&	0.177		\\
\end{tabular}		
\end{ruledtabular}		
\end{table*}		
		
\subsubsection{One-sided or two-sided?}

Because of the constraints of movement in 1D, electrons shield one another very effectively and, as a result, the outer electron lies far from the nucleus in the \ce{A$_{1,2}^{Z-2}$} state. 
Because of this, the \ce{A$_{1,2}^{Z-2}$} state is significantly higher in energy than the \ce{_1A$_1^{Z-2}$} state. 
For example, the HF energies of \ce{He$_{1,2}$} and \ce{_1He1} are $-2.107356$ and $-3.242922$, respectively.

In the hydride anion \ce{H^-} ($Z=1$), the nucleus cannot bind the second electron in the \ce{H$_{1,2}^-$} state and this species autoionizes. 
The corresponding state of the helium atom is bound but its ionization energy is only 0.1074. 
Whereas the minimum nuclear charge which can bind two electrons is $Z_\text{crit} \approx 1.1$ in the \ce{A$_{1,2}^{Z-2}$} state, it is $Z_\text{crit} \approx 0.65$ in the \ce{_1A$_1^{Z-2}$} state.
In comparison, Baker \textit{et al.}~have reported\cite{Baker90} that the corresponding value in 3D is $Z_\text{crit} \approx 0.91$.

In the \ce{_1A$_1^{Z-2}$} state, each electron is confined to one side of the nucleus, and is perfectly shielded from the other electron by the nucleus.
As a result, the electron correlation energy $\Ec$ is entirely of the dispersion type and is much smaller than in 3D atoms.
For example, $\Ec$ in \ce{_1He1} is $-3.022$ while $\Ec$ in the ground state of 3D He is $-42.024$. 
It is interesting to note that, unlike the situation in 3D, the correlation energy of \ce{_1H1-} is slightly larger than in \ce{_1He1} and approaches the large-$Z$ limit from below.

Table \ref{tab:Helike} also shows that correlation energies $\Esoft$ arising from use of the softened Coulomb operator\cite{Wagner12} are totally different from energies $\Ec$ from the unmodified operator.  This qualitative change arises because the softened operator allows the electrons to share the same orbital.

\subsubsection{Large-$Z$ expansion}

In the large-$Z$ (i.e.~high-density) limit, the exact and HF energies of the two-sided He-like ions can be expanded as a power series using Rayleigh-Schr{\"o}dinger perturbation theory \cite{EcLimit09}
\begin{align}
	\Eex	& = E^{(0)}\,Z^2 + E^{(1)}\,Z + E^{(2)} + \frac{E^{(3)}}{Z} + O(Z^{-2}),
	\\
	\EHF & =E_\text{HF}^{(0)}\,Z^2 + E_\text{HF}^{(1)}\,Z + E_\text{HF}^{(2)} + \frac{E_\text{HF}^{(3)}}{Z} + O(Z^{-2}),
\end{align}
where 
\begin{align}
	E^{(0)} & = E^{(0)}_\text{HF} = -1,
	&
	E^{(1)} & = E^{(1)}_\text{HF} = 2/5. 
\end{align}
For large $Z$, the limiting correlation energy is thus 
\begin{equation}
	\Ec 	= E^{(2)}  - E_\text{HF}^{(2)} + \frac{E^{(3)}-E_\text{HF}^{(3)}}{Z} + O(Z^{-2})
		= \Ec^{(2)} + \frac{\Ec^{(3)}}{Z} + O(Z^{-2}).
\end{equation}
The second- and third-order exact energies
\begin{align}
	E^{(2)} & =  -0.045545, 
	&
	E^{(3)} & = -0.000650,
\end{align}
can be found by Hylleraas' approach,\cite{Hylleraas30} while the second- and third-order HF energies
\begin{align}
	E_\text{HF}^{(2)} & = -0.042832,
	&
	E_\text{HF}^{(3)} & = -0.000495,
\end{align}
can be found by Linderberg's method. \cite{Linderberg60, Linderberg61} We conclude, therefore, that
\begin{equation}
	\Ec = -2.713 - \frac{0.155}{Z} + O(Z^{-2}).
\end{equation}
The negative sign of $\Ec^{(3)}$ explains the reduction in the correlation energy as $Z$ increases.

It is interesting to note that the 2D and 3D values of $\Ec^{(2)}$ are $-220.133$ and $-46.663$, respectively,\cite{EcLimit09, EcProof10, Frontiers10} which are much larger than the corresponding 1D values.

\subsection{Periodic Table}
\begin{table*}
\caption{\label{tab:TableAtom} Energies, HOMO-LUMO gaps, dipole moments and radii of 1D atoms and ions}
\small
\begin{ruledtabular}
\begin{tabular}{lcccccccc}
							&			\mc{2}{c}{Energy}			&	\mc{3}{c}{Correlation energy}		&		\mc{3}{c}{HF property}		\\
											\cline{2-3}								\cline{4-6}							\cline{7-9}		
Ion							&	$-\Eex$			&	$-\EHF$		&	$-\EP$	&	$-\EPP$	&	$-\Ec$	&	Gap	& 	\dip 	& 	\rms	\\
\hline
\ce{$\ $H+}				&	0				&	0				&	0		&	0			&	0		&	---		&	0		&	0		\\
\ce{$\ $H1}				&	0.500 000		&	0.500 000		&	0		&	0			&	0		&	0.373	&	1.500	&	1.732	\\
\ce{_1H1-}				&	0.646 584		&	0.643 050		&	1.715	&	2.530		&	3.534	&	0.168	&	0		&	2.296	\\[3mm]

\ce{$\ $He1+}				&	2.000 000		&	2.000 000		&	0		&	0			&	0		&	0.776	&	0.750	&	0.866	\\
\ce{_1He1}				&	3.245 944		&	3.242 922		&	2.063	&	2.688		&	3.022	&	1.264	&	0		&	0.985	\\
\ce{_1He$_{1,2}^-$}			&											\mc{8}{c}{Autoionizes}													\\[3mm]

\ce{_1Li1+}					&	7.845 792		&	7.842 889		&	2.235	&	2.733		&	2.903	&	3.200	&	0		&	0.628	\\
\ce{_1Li$_{1,2}$}			&	8.011 9\ \ \ \ 	&	8.007 756		&	3.36	&	4.03		&	4.1		&	0.119	&	1.483	&	2.836	\\
\ce{$_{1,2}$Li$_{1,2}^-$}	&					&	8.059 016		&	3.92	&	4.75		&			&	0.062	&	0		&	4.219	\\[3mm]

\ce{$_1$Be$_{1,2}^+$}		&	15.041 1\ \ \ \ \ 	&	15.035 639		&	4.77	&	5.48		&	5.5		&	0.220	&	0.829	&	1.599	\\
\ce{$_{1,2}$Be$_{1,2}$}	&					&	15.415 912		&	6.68	&	7.69		&			&	0.386	&	0		&	2.111	\\
\ce{$_{1,2}$Be$_{1-3}^-$}	&											\mc{8}{c}{Autoionizes}													\\[3mm]

\ce{$_{1,2}$B$_{1,2}^+$}	&					&	25.281 504		&	8.75	&	9.80		&			&	0.897	&	0		&	1.437	\\
\ce{$_{1,2}$B$_{1-3}$}		&					&	25.357 510		&	9.7		&	10.9		&			&	0.056	&	1.881	&	4.655	\\
\ce{$_{1-3}$B$_{1-3}^-$}	&					&	25.380 955		&	9.97	&	11.33		&			&	0.036	&	0		&	7.042	\\[3mm]

\ce{$_{1,2}$C$_{1-3}^+$}	&					&	37.918 751		&	12.8	&	14.3		&			&	0.104	&	1.070	&	2.726	\\
\ce{$_{1-3}$C$_{1-3}$}		&					&	38.090 383		&	14.6	&	16.5		&			&	0.176	&	0		&	3.684	\\
\ce{$_{1-3}$C$_{1-4}^-$}	&											\mc{8}{c}{Autoionizes}													\\[3mm]

\ce{$_{1-3}$N$_{1-3}^+$}	&					&	53.528 203		&	18.7	&	20.9		&			&	0.400	&	0		&	2.557	\\
\ce{$_{1-3}$N$_{1-4}$}		&					&	53.569 533		&	19.1	&	21.5		&			&	0.031	&	2.423	&	7.139	\\
\ce{$_{1-4}$N$_{1-4}^-$}	&					&	53.582 040		&	19.3	&	21.7		&			&	0.030	&	0		&	11.094	\\[3mm]

\ce{$_{1-3}$O$_{1-4}^+$}	&					&	71.836 884		&	23.8	&	26.6		&			&	0.059	&	1.382	&	4.267	\\
\ce{$_{1-4}$O$_{1-4}$}		&					&	71.929 302		&	24.9	&	28.1		&			&	0.098	&	0		&	5.806	\\
\ce{$_{1-4}$O$_{1-5}^-$}	&											\mc{8}{c}{Autoionizes}													\\[3mm]

\ce{$_{1-4}$F$_{1-4}^+$}	&					&	93.125 365		&	30.5	&	34.2		&			&	0.217	&	0		&	4.048	\\
\ce{$_{1-4}$F$_{1-5}$}		&					&	93.149 851		&	30.7	&	34.5		&			&	0.020	&	2.939	&	10.041	\\
\ce{$_{1-5}$F$_{1-5}^-$}	&					&	93.157 319		&	31		&	35			&			&	0.037	&	0		&	15.538	\\[3mm]

\ce{$_{1-4}$Ne$_{1-5}^+$}	&					&	117.256 746	&	36.3	&	40.9		&			&	0.037	&	1.745	&	6.246	\\
\ce{$_{1-5}$Ne$_{1-5}$}	&					&	117.312 529	&	37		&	42			&			&	0.067	&	0		&	8.586	\\
\ce{$_{1-5}$Ne$_{1-6}^-$}	&											\mc{8}{c}{Autoionizes}													\\
\end{tabular}		
\end{ruledtabular}		
\end{table*}		

We have computed the ground-state energies of the 1D atoms from \ce{Li} to \ce{Ne} at the HF, MP2 and MP3 levels. 
We have also computed these energies for their cations and anions.
To compute the exact energy of \ce{Li} and \ce{Be+}, we have used a Hylleraas-type wave function containing a large number of terms.
The results are reported in Table \ref{tab:TableAtom}. 

Where exact energies are available, it appears that the MP2 and MP3 calculations recover a large proportion of the exact correlation energy.
Their performance appears to improve rapidly as the atomic number grows and, for this reason, we consider the MP3 energies to be reliable benchmarks for the heavy atoms.

In view of the modest sizes of these atomic correlation energies, we conclude that it is likely that, for 1D systems, even the simple HF model is reasonably accurate and MP2 offers a very accurate theoretical model chemistry.

The accuracy of perturbative methods throughout Table \ref{tab:TableAtom} may be surprising given the small band gaps in some of the species, e.g.~\ce{Li}. 
Although a small gap is often an indicator of poor performance for perturbative corrections, the associated HOMO-LUMO excitations correspond to the movement of an electron from the outermost orbital on one side of the nucleus to the corresponding orbital on the other side, e.g.~exciting from \ce{_1Li_{1,2}} to \ce{$_{1,2}$Li1}.  However, such excitations are excluded from the perturbation sums because they involve the (physically forbidden) movement of an electron from one domain to another. 

\begin{table*}
	\caption{\label{tab:IEsEAs} Ionization energy and electron affinity (in eV) of 1D atoms}
	\small
	\begin{ruledtabular}
		\begin{tabular}{lcccccc}
						&		\mc{3}{c}{Ionization energy}		&	\mc{3}{c}{Electron affinity}		\\
			Atom		&		\mc{3}{c}{\ce{A -> A+ + e-} }			&	\mc{3}{c}{\ce{A + e- -> A-}}		\\
									\cline{2-4}									\cline{5-7}
						&	HF		&	MP2		&	MP3		&	HF		&	MP2	&	MP3	\\
			\hline
			\ce{H}		&	13.606	&	13.606		&	13.606		&	3.893	&	3.939	&	3.961	\\
			\ce{He}		&	33.822	&	33.878		&	33.895		&	0		&	0		&	0		\\
			\ce{Li}		&	4.486	&	4.517		&	4.522		&	1.395	&	1.410	&	1.414	\\
			\ce{Be}		&	10.348	&	10.400		&	10.408		&	0		&	0		&	0		\\
			\ce{B}		&	2.068	&	2.09		&	2.098		&	0.643	&	0.651	&	0.655	\\
			\ce{C}		&	4.670	&	4.719		&	4.733		&	0		&	0		&	0		\\
			\ce{N}		&	1.125	&	1.14		&	1.14		&	0.340	&	0.35	&	0.35	\\
			\ce{O}		&	2.515	&	2.54		&	2.56		&	0		&	0		&	0		\\
			\ce{F}		&	0.666	&	0.67		&	0.67		&	0.203	&	0.21	&	0.2		\\
			\ce{Ne}		&	1.518	&	1.5			&	1.5			&	0		&	0		&	0		\\
		\end{tabular}		
	\end{ruledtabular}		
\end{table*}		

We have computed the ionization energy IE (\ce{A -> A+ + e-}) and the electron affinity EA (\ce{A + e- -> A-}) of each atom and these are summarised in Table \ref{tab:IEsEAs}.  Our HF calculations revealed that anions of even-$Z$ atoms (viz.~\ce{He-}, \ce{Be-}, \ce{C-}, \ce{O-} and \ce{Ne-}) autoionize. 
The IEs display a clear zig-zag pattern as the atomic number grows, reminiscent of the IEs in 3D.  However, in 1D the period is very short, viz.~two.

\begin{figure}
	\includegraphics[width=0.2\textwidth]{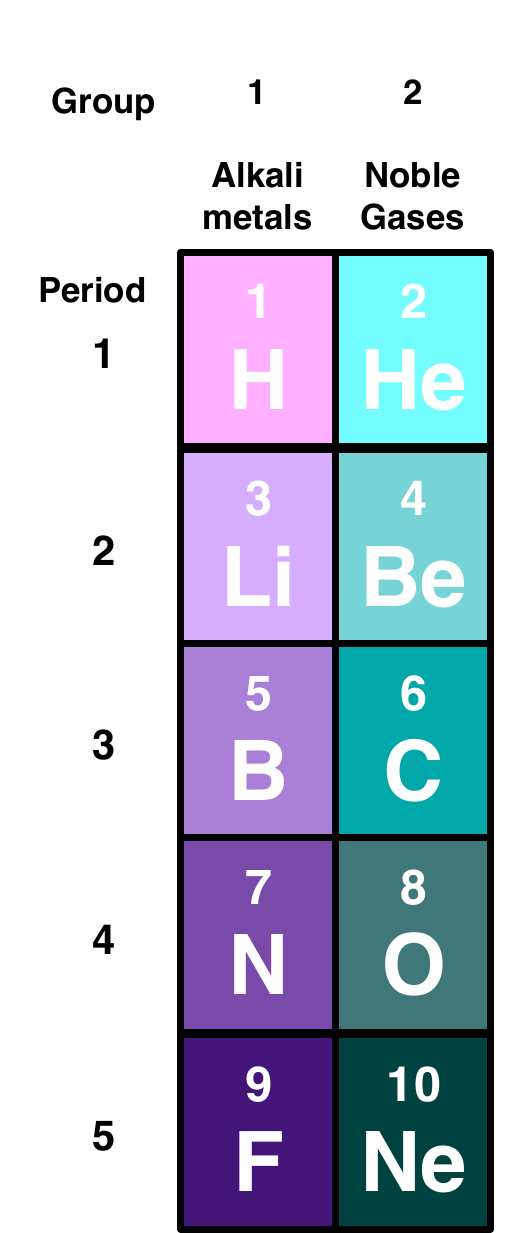}
	\caption{
	\label{fig:PeriodicTable}
	Periodic table in 1D}
\end{figure}

The odd-$Z$ atoms have a non-zero dipole moment, which allows reactivity with other odd-$Z$ atoms via dipole-dipole interactions.  In contrast, the even-$Z$ atoms have only a quadrupole and would be expected to be more electrostatically inert. 
The combination of the periodic trends in the IEs and the pattern of atomic reactivities allows us to construct a periodic table for 1D atoms (Fig. \ref{fig:PeriodicTable}). The 1D atoms \ce{H}, \ce{Li}, \ce{B}, \ce{N} and \ce{F} are the analogs of the 3D alkali metals (i.e. \ce{H}, \ce{Li}, \ce{Na}, \ce{K} and \ce{Rb}) and the 1D atoms \ce{He}, \ce{Be}, \ce{C}, \ce{O} and \ce{Ne} are the analogs of the 3D noble gases (i.e. \ce{He}, \ce{Ne}, \ce{Ar}, \ce{Kr} and \ce{Xe}).

Like their 3D analogs,\cite{Lykke91, Haeffler96, Hotop85, Andersson00, Frey78, HandbookChemPhys} the 1D IEs drop as the nuclear charge increases. However, this behaviour is more dramatic in 1D than in 3D because the strong shielding in 1D causes the outermost electrons to be very weakly attracted to the nucleus.  This effect is so powerful that the third 1D noble gas (\ce{C}) has an IE (4.733 eV) which is lower than the IE (5.139 eV) of the third 3D alkali metal (\ce{Na}).

1D EAs also behave similarly to their 3D counterparts, decreasing as the nuclear charge increases. 
Because one side of the nucleus is completely unshielded, the EA of 1D \ce{H} (3.961 eV) is far larger than that of 3D \ce{H} (0.754 eV). 
However, like the 1D IEs, shielding effects lead to a rapid reduction in EA as the nuclear charge increases. 
As a result, the fifth 1D alkali metal (\ce{F}) has an EA  (0.160 eV) which is considerably smaller than the EA (0.486 eV) of the fifth 3D alkali metal (\ce{Rb}).

We have also computed \rms\ as a measure of atomic radius and compared these to the calculated values of Clementi \textit{et al.}\cite{Clementi63, Clementi67} for 3D atoms.  Whereas a 3D alkali metal atom is much larger than the noble gas atom of the same period, the 1D alkali metal atoms are only slightly larger than their noble gas counterparts.

\section{
\label{sec:molecules}
Molecules}

\subsection{
\label{sec:1e-diatomics}
One-electron diatomics}
\begin{table*}
\caption{
\label{tab:TableMol}
Structures, energies, gaps and vibrational frequencies $\nu$ (in cm$^{-1}$) of 1D molecules}
\footnotesize
\begin{ruledtabular}
\begin{tabular}{lllccccccc}
				&	\mc{2}{c}{Molecule}						&	\mc{2}{c}{Total energy}		&	\mc{3}{c}{Correlation energy}	&	Gap	& 	$\nu$ 	\\
\cline{2-3}																\cline{4-5}						\cline{6-8}
				&	State			&	Bond length		&	$-\Eex$		&	$-\EHF$	&$-\EP$	&$-\EPP$	&$-\Ec$							\\
\hline
				&  \ce{H1H+}		& 	$\Req = 2.581$		&	0.830 710	&	0.830 710	&	0		&	0		&	0		&	3.42	&	2470	\\
One-electron 	&  \ce{He1H^2+}	& 	$\Req = 2.182$		&	1.830 303	&	1.830 303	&	0		&	0		&	0		&	2.39	&	3553	\\
diatomics		&					& 	$\Rts = 3.296$		&	1.809 411	&	1.809 411	&	0		&	0		&	0		&	1.28	&	1914	\\
				&  \ce{He1He^3+}	& 	$\Req = 1.793$		&	1.986 928	&	1.986 928	&	0		&	0		&	0		&	9.89	&	4267	\\
				&					& 	$\Rts = 4.630$		&	1.694 543	&	1.694 543	&	0		&	0		&	0		&	1.48	&	1028	\\
\hline
				&  \ce{H1H1}		&	$\Req = 2.639$		&	1.185 948	&	1.184 571	&	1.400	&	1.374	&	1.377	&	0.264	&	2389	\\
Two-electron	&  \ce{_1He1H+}	&	$\Req = 2.016$		&	3.444 390	&	3.441 957	&	2.457	&	2.438	&	2.433	&	1.220	&	3747	\\
diatomics		&  \ce{He1H1+}		&	$\Req = 2.037$		&	2.517 481	&	2.516 810	&	0.681	&	0.669	&	0.671	&	0.443	&	3939	\\
				&  \ce{He1He1^2+}	&	$\Req = 1.668$		&	4.112 551	&	4.110 780	&	1.784	&	1.772	&	1.771	&	1.480	&	4755	\\
				&					&	$\Rts = 3.989$		&	3.807 432	&	3.807 165	&	0.251	&	0.259	&	0.267	&	0.307	&	1286	\\
\hline
Triatomics		&  \ce{H1H1H+}	& 	$\Req = 2.664$		&	1.570 720	&	1.569 820	&	0.918	&	0.897	&	0.900	&	1.557
																															&	1178\footnotemark[1]	\\
\end{tabular}		
\end{ruledtabular}
\begin{flushleft}
$^\text{a}$Symmetric vibrational mode.
\end{flushleft}
\end{table*}	

The electronic Hamiltonian of a one-electron diatomic \ce{AB^{$Z_\text{A}$+$Z_\text{B}$-1}} composed of two nuclei A and B of charges $Z_\text{A}$ and $Z_\text{B}$ located at $x = -R/2$ and $x = +R/2$ is
\begin{equation}
\label{H-1e-diatomics}
	\Hat{H} = -\frac{1}{2} \frac{d^2}{dx^2} - \frac{Z_\text{A}}{|x+R/2|} - \frac{Z_\text{B}}{|x-R/2|}.
\end{equation}
For these systems, three families of states are of interest:
\begin{itemize}
	\item The \ce{$_i$AB$^{Z_\text{A}+Z_\text{B}-1}$} and \ce{AB$_i^{Z_\text{A}+Z_\text{B}-1}$} families where the electron is outside the nuclei;\footnotemark[2]
	\footnotetext[2]{In the homonuclear case, i.e. $Z_\text{A} = Z_\text{B}$, these two families are equivalent.}
	\item The \ce{A$_i$B$^{Z_\text{A}+Z_\text{B}-1}$} family where the electron is between the two nuclei.
\end{itemize}
Some of the properties of three such systems are reported in the upper half of Table \ref{tab:TableMol}.

\begin{figure}
	\includegraphics[height=0.3\textheight]{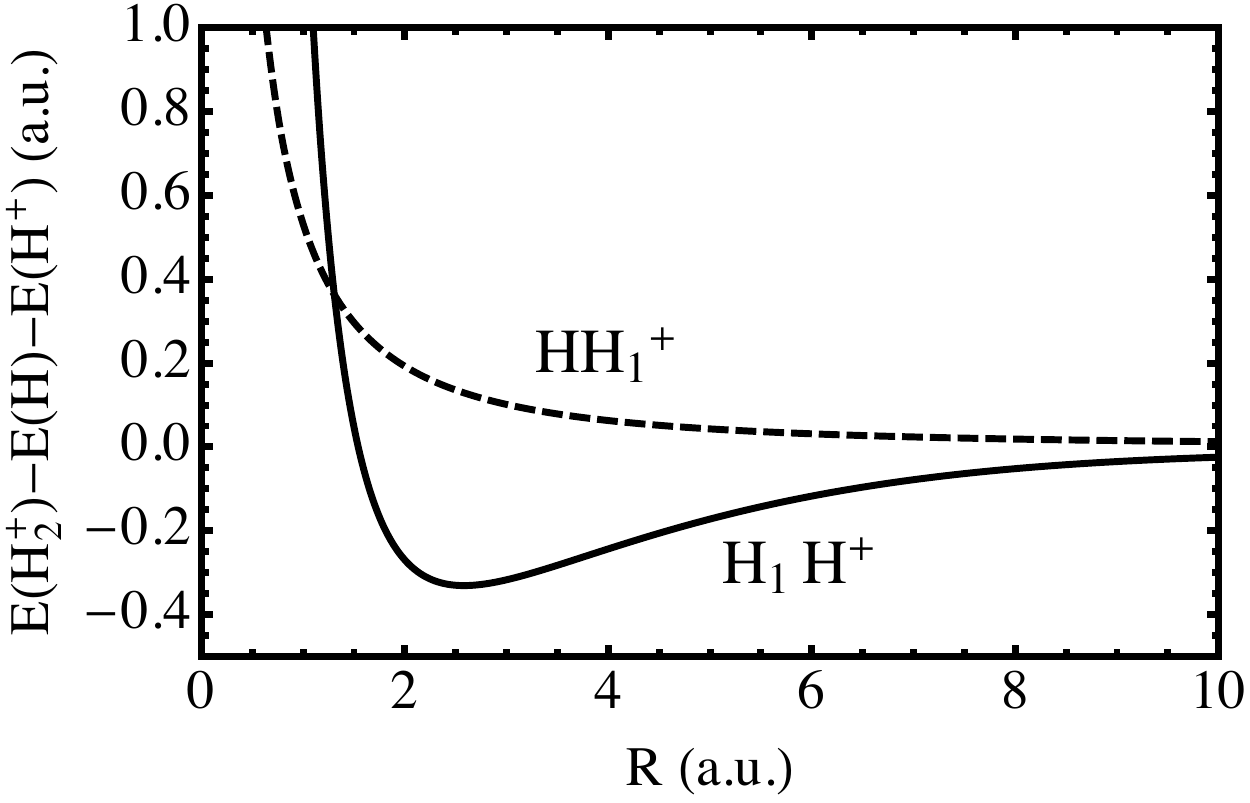}
	\caption{
	\label{fig:H2plus}
	Potential energy curves of the \ce{H1H+} and \ce{HH1+} states of \ce{H2+}}
\end{figure}

\subsubsection{\ce{H2+}}

The simplest of all molecules is the homonuclear diatomic \ce{H2+}, in which $Z_\text{A} = Z_\text{B} = 1$. 
In 3D, this molecule was first studied by Burrau who pointed out that the Schr\"odinger equation is separable in confocal elliptic coordinates. \cite{Burrau27}
In 1928, Linus Pauling published a review summarizing the work of Burrau and many other researchers.\cite{Pauling28, PaulingBook}
In Appendix \ref{app:exact}, we report some exact wave functions for \ce{H1H+} in 1D.

The near-exact potential energy curves of the \ce{H1H+} and \ce{HH1+} states are shown in Fig.~\ref{fig:H2plus}.
Beyond $R = 1.5$, the \ce{H1H+} state is lower in energy than the \ce{HH1+} state.  However, when the bond is compressed, the kinetic energy of the trapped electron becomes so large that the \ce{H1H+} state rises above the \ce{HH1+} state.  
The bond dissociation energy (0.3307 $\Eh$) of \ce{H1H+} is large and its equilibrium bond length ($R_\text{eq} = 2.581$ bohr) is long. 
Both values are much larger than the corresponding 3D values (0.1026 $\Eh$ and 1.997 bohr).\cite{Scott06}
Whereas the \ce{H1H+} state is bound by a favorable charge-dipole interaction, the \ce{HH1+} state is repulsive because of a similar, but unfavorable, interaction.
Using this simple electrostatic argument, one can predict that the \ce{H1H+} and \ce{HH1+} potential energy curves behave as $-\mu_{\ce{H}}/R^2$ and $+\mu_{\ce{H}} /R^2$ for large $R$, where $\mu_\ce{H} = 3/2$.  This charge-dipole model is qualitatively correct for $R \gtrsim 10$ for \ce{H1H+} and $R \gtrsim 5$ for \ce{HH1+}.

\begin{figure}[htbp]
	\includegraphics[height=0.29\textheight]{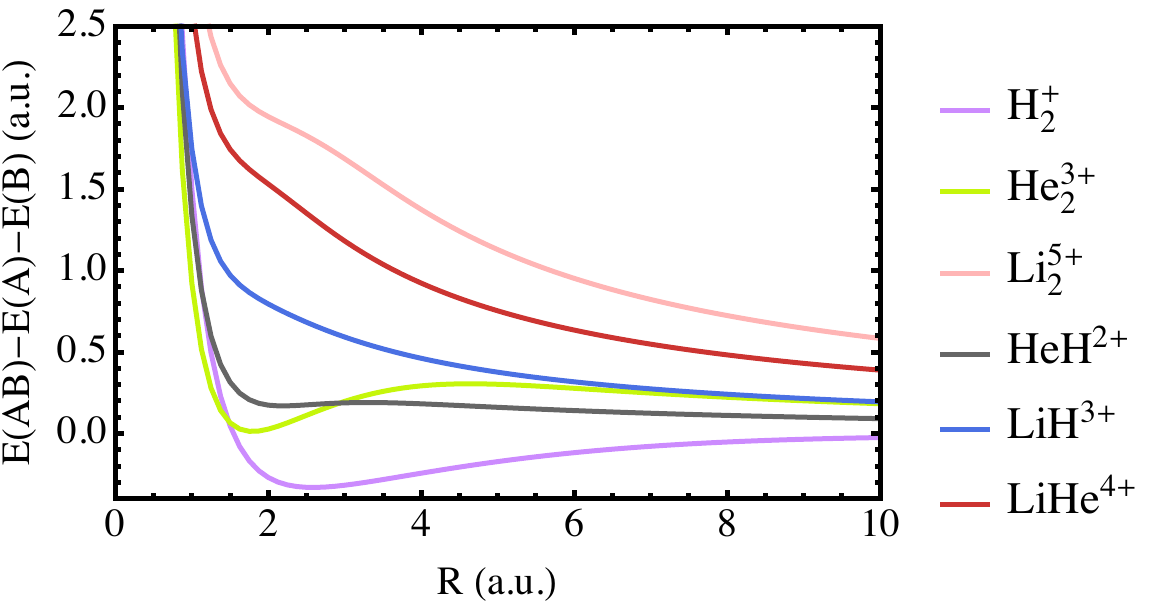}
	\caption{
	\label{fig:1e-molecule}
	Potential energy curves of the \ce{A1B$^{Z_\text{A}+Z_\text{B}-1}$} states of several one-electron diatomics}
\end{figure}

\subsubsection{\ce{HeH^2+} and \ce{He2^3+}}

The Hamiltonians of \ce{HeH^2+} and \ce{He2^3+} are given by \eqref{H-1e-diatomics} for $Z_\text{A} = 1$ and $Z_\text{B} = 2$, and  $Z_\text{A} = Z_\text{B} = 2$, respectively.  As in \ce{H2+}, we find that \ce{He1He^3+} is more stable than \ce{HeHe1^3+}, and \ce{He1H^2+} is more stable than \ce{HeH1^2+} and \ce{_1HeH^{2+}}, except at short bond lengths.

In 3D, the molecules \ce{HeH^2+} and \ce{He2^3+} are unstable except in strong magnetic fields. \cite{Turbiner10} However, as Fig.~\ref{fig:1e-molecule} shows, \ce{He1H^2+} and \ce{He1He^3+} are metastable species in 1D with equilibrium bond lengths of $R_\text{eq} = 2.182$ and 1.793, and transition structure bond lengths of $R_\text{ts} =3.296$ and 4.630, respectively.  Although these species are thermodynamically unstable with respect to \ce{He+ + H+} and \ce{He+ + He^2+}, they are protected from dissociation by barriers of 0.0209 and 0.2924, respectively.  For large $R$, their dissociation curves behave as $1/R - \mu_\ce{He+}/R^2$ and $2/R - 2\mu_\ce{He+}/R^2$, respectively, where $\mu_\ce{He+} = 3/4$.

All the heavier one-electron diatomics have purely repulsive dissociation curves.

\subsubsection{Chemical bonding in one-electron diatomics}

\begin{figure}
	\includegraphics[height=0.3\textheight]{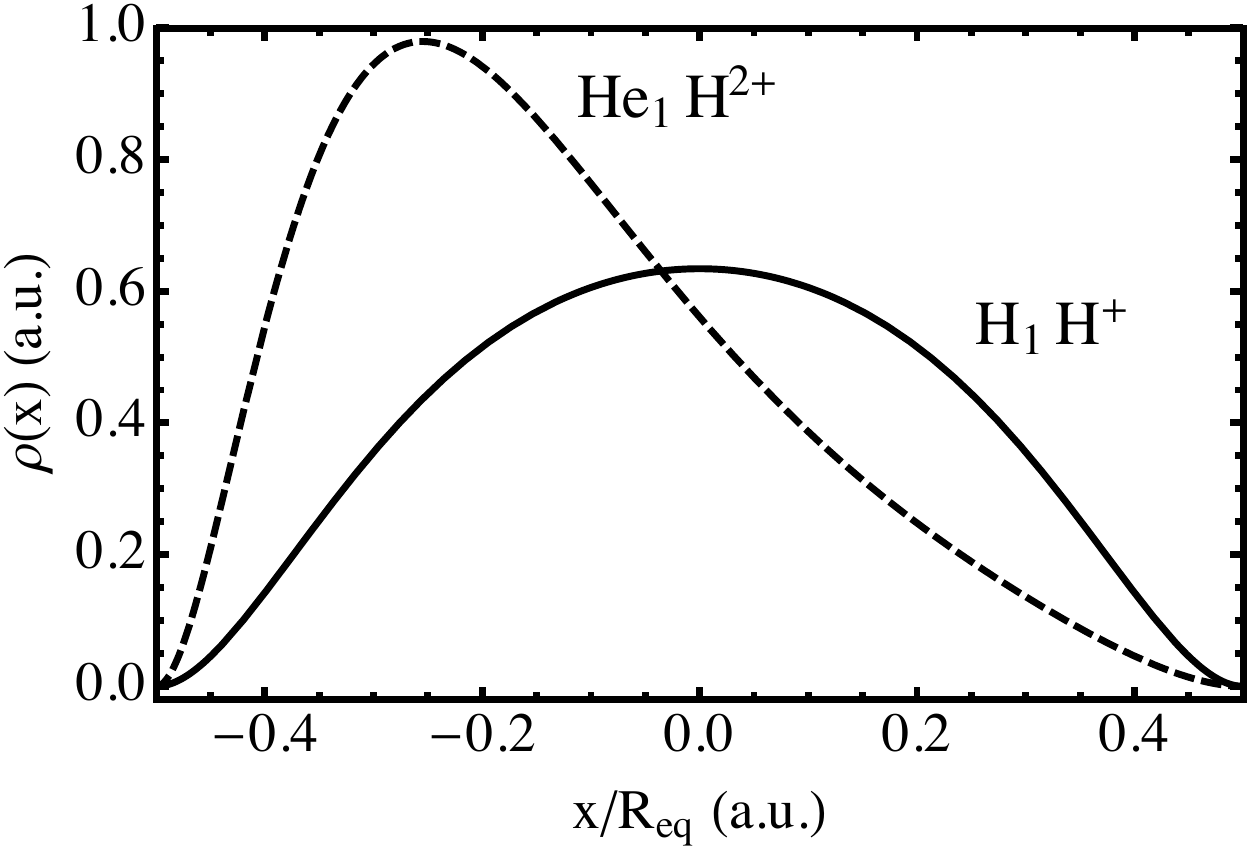}
	\caption{
	\label{fig:H2plus-bond}
	Electronic density $\rho(x)$ in \ce{H1H+} and \ce{He1H^2+} at their equilibrium bond lengths}
\end{figure}

Fig.~\ref{fig:H2plus-bond} shows the electronic density $\rho(x)$ for \ce{H1H+} and \ce{He1H^2+} at their equilibrium bond lengths. 
Whereas the electron density in a typical 3D bond is greatest at the nuclei and reaches a minimum near the middle of the bond,\cite{PaulingBook} the electron density in these 1D bonds vanishes at the nuclei and achieves a maximum in the middle of the bond.
The bond in \ce{He1H^2+} is polarized towards the nucleus with the largest charge. 

\subsubsection{Harmonic vibrations}

We have computed the harmonic vibrational frequencies of \ce{H1H+}, \ce{He1H^2+} and \ce{He1He^3+} at their equilibrium bond lengths and these are shown in Table \ref{tab:TableMol}.
The second derivative of the energy was obtained numerically using the three-point central difference formula and a stepsize of $10^{-2}$ bohr.
The frequency of the 1D \ce{H1H+} ion (2470 cm$^{-1}$) is similar to that of the 3D ion (2321 cm$^{-1}$)\cite{HerzbergBook} but this result is probably accidental.
Although the barrier in \ce{He1H^2+} is small and its harmonic frequency relatively high (3553 cm$^{-1}$), the ion probably supports a vibrational state: the zero-point vibrational energy is only half the barrier height.

\subsection{Two-electron diatomics}
The Hamiltonian of a two-electron diatomic \ce{AB^{$Z_\text{A}$+$Z_\text{B}$-2}} composed of two nuclei A and B of charges $Z_\text{A}$ and $Z_\text{B}$ located at $x = -R/2$ and $x = +R/2$ is
\begin{equation}
\label{H-2e-diatomics}
	\Hat{H} = -\frac{1}{2} \left( \frac{\partial^2}{\partial x_1^2}+\frac{\partial^2}{\partial x_2^2} \right) - \frac{Z_\text{A}}{\left|x_1+\frac{R}{2}\right|} 
	- \frac{Z_\text{A}}{\left|x_2+\frac{R}{2}\right|} - \frac{Z_\text{B}}{\left|x_1-\frac{R}{2}\right|} - \frac{Z_\text{B}}{\left|x_2-\frac{R}{2}\right|} + \frac{1}{|x_1 - x_2|}.
\end{equation}
These systems possess six families of states:
\begin{itemize}
	\item The \ce{A$_{i,j}$B$^{Z_\text{A}+Z_\text{B}-2}$} family;
	\item The \ce{$_i$AB$_j^{Z_\text{A}+Z_\text{B}-2}$} family;
	\item The \ce{$_i$A$_j$B$^{Z_\text{A}+Z_\text{B}-2}$} and \ce{A$_i$B$_j^{Z_\text{A}+Z_\text{B}-2}$} families; \footnotemark[2]
	\item The \ce{$_{i,j}$AB$^{Z_\text{A}+Z_\text{B}-2}$} and \ce{AB$_{i,j}^{Z_\text{A}+Z_\text{B}-2}$} families; \footnotemark[2]
\end{itemize}
Some of the properties of four such systems are reported in the lower half of Table \ref{tab:TableMol}. 

\subsubsection{\ce{H2}}

\begin{figure}	
	\includegraphics[height=0.3\textheight]{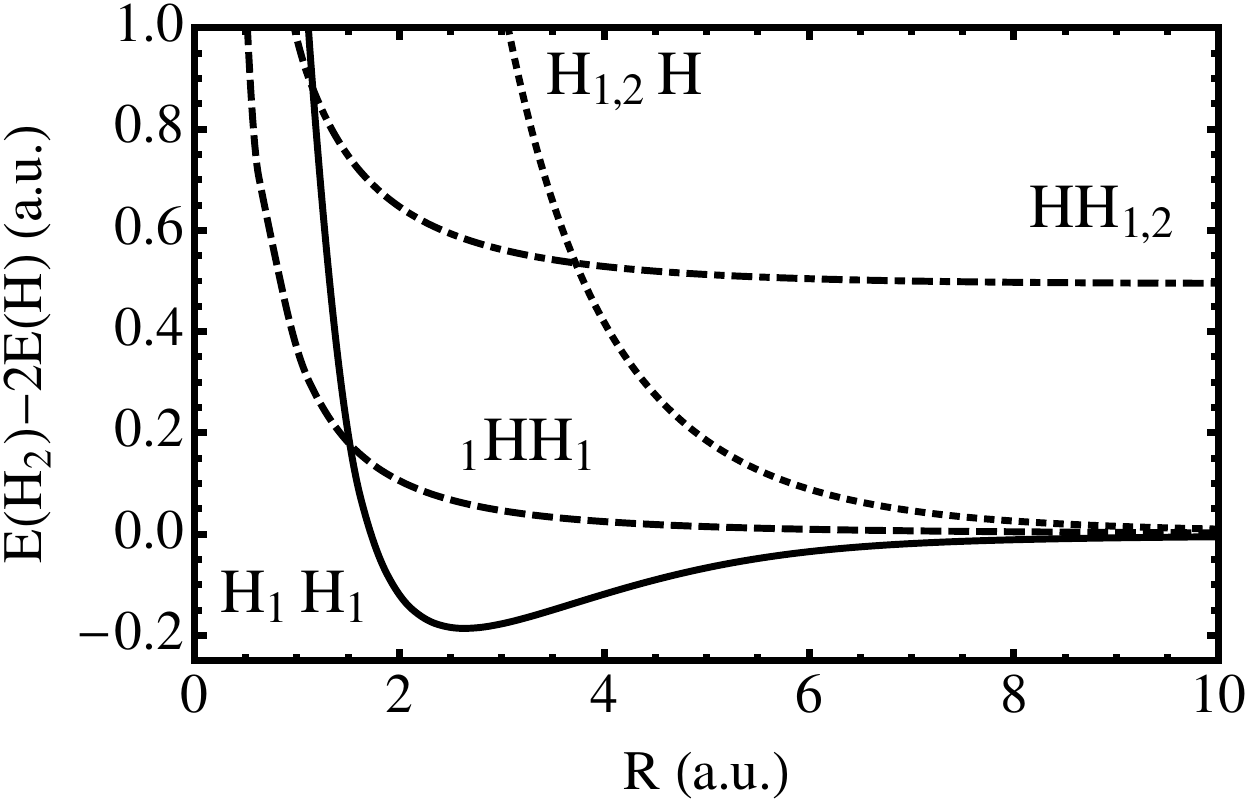}
	\caption{
	\label{fig:H2}
	Potential energy curves of the \ce{H1H1}, \ce{_1HH1}, \ce{H$_{1,2}$H} and \ce{HH$_{1,2}$} states of \ce{H2}}
\end{figure}

The simplest two-electron diatomic is \ce{H2} where $Z_\text{A} = Z_\text{B} = 1$. 
The 3D version of this molecule has been widely studied since the first accurate calculation of James and Coolidge \cite{James33} in 1933. 
The ground state in each family has been calculated using Hylleraas-type calculations and is represented in Fig.~\ref{fig:H2}.  We note that the HF and Hylleraas curves are almost indistinguishable due to the small correlation energy in these systems (see Table \ref{tab:TableMol}).

As expected, \ce{HH$_{1,2}$} is high in energy due to shielding by the inner electron (see discussion on the He-like ions in Sec.~\ref{sec:Helike}), and dissociates into \ce{H+ + H_{1,2}^-}. 
The three other states dissociate into a pair of H atoms.
As in \ce{H2+}, the \ce{_1HH1} state is the most stable at small bond lengths, but is higher in energy than \ce{H1H1} when $R > 1.5$ bohr. 
The \ce{H1H1} state is bound with an equilibrium bond length of $2.639$ bohr and a dissociation energy of $0.1859$ $\Eh$.
In comparison, the bond length of the 3D \ce{H2} molecule is close to $1.4$ bohr and has a similar dissociation energy ($0.1745$ $\Eh$). \cite{Kolos68}
The harmonic vibrational frequency of \ce{H1H1} (2389 cm$^{-1}$) is significantly lower than the 3D value (4401 cm$^{-1}$). \cite{HerzbergBook} 
The equilibrium bond lengths and vibrational frequencies of \ce{H1H+} and \ce{H1H1} are similar because of the efficient shielding in 1D. 
Finally, we note that \ce{H1H1} has a non-zero dipole moment and the two fragments \ce{H1} are bound by a dipole-dipole interaction.

For those who are familiar with the traditional covalent two-electron bond in 3D chemistry, the instability of \ce{H$_{1,2}$H} is probably surprising.
However, this state is destabilized by two important effects: (a) the high kinetic energy of the electrons when trapped between nuclei (see discussion on \ce{H2+} in Sec.~\ref{sec:1e-diatomics}) and (b) the 1D exclusion principle, which mandates that the second electron occupy a higher-energy orbital than the first.  For these reasons, 1D molecules are usually held together by one-electron bonds (sometimes called hemi-bonds).

Bonding in \ce{H2+}, which is driven by the \ce{H+ + H} charge-dipole interaction, is roughly twice as strong as the bonding in \ce{H2}, which arises from the much weaker \ce{H + H} dipole-dipole interaction.  In contrast, in 3D, the \ce{H2} bond is roughly twice as strong as that in \ce{H2+}.

We expect that two-electron (or more) bonds exist in neutral species such as \ce{$_1$Li$_{1,2}$H$_1$} because of favorable dipole-dipole interactions.  However, such species are bound \emph{despite} the two-electron bond, rather than because of it, and are probably very weakly bound.  We plan to investigate this further in a forthcoming paper. \cite{HF1D}

\subsubsection{\ce{HeH+} and \ce{He2^2+}}

The Hamiltonian for \ce{HeH+} and \ce{He2^2+} are given by \eqref{H-2e-diatomics} for $Z_\text{A} = 1$ and $Z_\text{B} = 2$, and $Z_\text{A} = Z_\text{B} = 2$, respectively. 
Like \ce{He1He^3+}, \ce{He1He1^2+} is metastable with a large energy barrier of 0.3051 $\Eh$ and a late transition structure with $\Rts/\Req \approx 2.5$. In 3D, the \ce{He2^2+} dication is also metastable but with an earlier transition structure ($\Rts/\Req \approx 1.5$). \cite{Pauling33, Guilhaus84, Gill87, Gill88a, Gill88b}

Like the 3D \ce{HeH+} molecule, \cite{Wolniewicz65} the 1D \ce{_1He1H^+} and \ce{He1H1^+} ions are bound. 
The dissociation of \ce{_1He1H+} into \ce{_1He1 + H+} requires 0.1981 $\Eh$ and is much more endothermic than the dissociation of \ce{He1H1+} into \ce{He1+ + H1}, which requires only 0.0174 $\Eh$.  Surprisingly, however, they have similar bond lengths and harmonic frequencies.

\begin{figure}[htbp]
	\includegraphics[height=0.3\textheight]{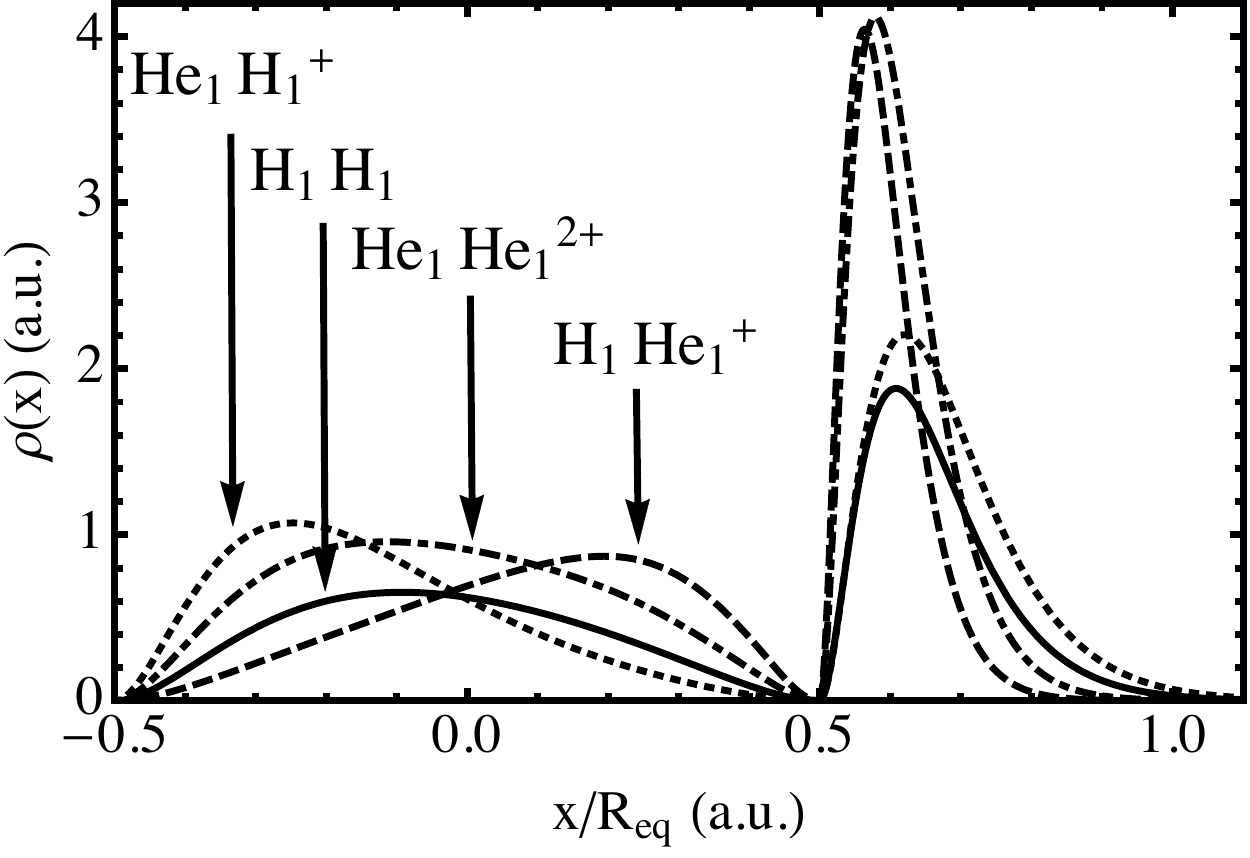}
	\caption{
	\label{fig:H2-bond}
	Electronic density $\rho(x)$ in \ce{H1H1}, \ce{H1He1+}, \ce{He1H1+} and \ce{He1He1^2+} at their equilibrium bond lengths}
\end{figure}

\subsubsection{Chemical bonding in two-electron diatomics}

Fig.~\ref{fig:H2-bond} shows the electronic densities $\rho(x)$ in \ce{H1H1}, \ce{H1He1+}, \ce{He1H1+} and \ce{He1He1^2+} at their respective equilibrium bond lengths.
The bonds in \ce{H1H1} and \ce{He1He1^{2+}} are polar because of the repulsion by the external electron.
In \ce{He1H1+}, the bond is highly polar because the repulsion by the external electron and the attraction of the He nucleus push in the same direction. 
In \ce{H1He1+}, the bond is polarized in the opposite direction because the repulsion by the external electron is dominated by the attraction of the He nucleus.

\subsubsection{Correlation effects}

Table \ref{tab:TableMol} reports the MP2, MP3 and exact correlation energies at the equilibrium geometries of \ce{H1H1}, \ce{_1He1H+}, \ce{He1H1+} and \ce{He1He1^2+}. 
All these values are small compared to their 3D analogs because correlation energy in these 1D systems is entirely due to dispersion.  As a result, correlation effects are pleasingly small and, for example, the HF bond length in \ce{H1H1} differs from the exact value by only $0.003$ bohr.
This re-emphasizes that the HF approximation is probably significantly more accurate in 1D than in 3D.

The range of $\Ec$ values ($-2.434$ in \ce{H1He1+}, $-1.771$ in \ce{He1He1^2+}, $-1.377$ in \ce{H1H1}, $-0.671$ in \ce{He1H1+}) can be rationalized by comparing the distance between the two electrons in each system (see Fig.~\ref{fig:H2-bond}):  shorter distances yield larger correlation energies.

For the diatomics in Table \ref{tab:TableMol}, HF theory is again found to be accurate and the MPn series appears to converge rapidly towards the exact correlation energies.  In particular, the MP3 and exact energies differ by only a few microhartrees.

\subsection{
\label{sec:H3}
Chemistry of \ce{H3+}}
The 3D \ce{H3+} ion was discovered by Thomson\cite{Thomson11} in 1911 and plays a central role in interstellar chemistry. \cite{Herbst00, Oka06, Oka13}
In astrochemistry, the main pathway for its production is 
\begin{equation}
\label{H3-3D}
	\ce{H2+ + H2 -> H3+ + H}
\end{equation}
and this reaction is highly exothermic ($\Delta U = -0.0639$ $\Eh$). \cite{Oka13}
In 3D, the ion has a triangular structure \cite{Mentch84} as first demonstrated by Coulson. \cite{Coulson35}
(See Ref.~\onlinecite{Oka13} for an interesting historical discussion on \ce{H3+}.)
The proton affinity of \ce{H2}
\begin{equation}
	\label{stab-H3-3D}
	\ce{H2 + H+ -> H3+} 
\end{equation}
is also strongly exothermic ($\Delta U = -0.1613$ $\Eh$). \cite{Carney76}

In this Section, we study the 1D analogs of these two reactions, viz.
\begin{align}
	\label{H3-1D}
	\ce{H1H+ + _1H1H} & \ce{->} \ce{H1H1H+ + _1H} 
	\\
	\label{stab-H3-1D}
	\ce{H1H1 + H+} & \ce{->} \ce{H1H1H+}
\end{align}
In 1D, the equilibrium structure of \ce{H1H1H+} has $D_{\infty h}$ symmetry, a bond length of 2.664 bohr, and an energy of $-1.570 720$ $\Eh$ (see Table \ref{tab:TableMol}). 
The correlation energy at this bond length is only $0.900$ $\mEh$.
Our calculations predict that reactions \eqref{H3-1D} and \eqref{stab-H3-1D} are both exothermic ($\Delta U = -0.0541$ and $-0.3848$ $\Eh$, respectively) and that reaction \eqref{stab-H3-1D} is barrierless.
It is interesting that the exothermicities of reactions \eqref{H3-3D} and \eqref{H3-1D} are close, and that the proton affinities (reactions \eqref{stab-H3-3D} and \eqref{stab-H3-1D}) are also broadly similar.

\subsection{
\label{sec:nanowire}
Hydrogen nanowire
}
Despite the fact that equi-spaced infinite \ce{H} chain in 3D suffers from a Peierls instability, \cite{Kertesz76} this system has attracted considerable interest due to its strong correlation character and metal-insulator transition. \cite{Hachmann06, Tsuchimochi09, Sinitskiy10, Stella11, Zgid11} 
We have therefore used periodic HF calculations\cite{Saunders84, Saunders94} to compute the energy per atom of an infinite chain of equi-spaced 1D \ce{H} atoms separated by a distance $R$. 
Motivated by our results for 1D \ce{H2+}, \ce{H2} and \ce{H3+}, we have studied the state in which one electron is trapped between each pair of nuclei, i.e. \ce{$\cdots$ H1H1H1H1 $\cdots$}.

We have expanded the HF orbital in the unit cell ($x \in [-R/2,R/2]$) as a linear combination of $K$ even polynomials \eqref{E}.
We find that, near the minimum-energy structure, $K=4$ suffices to achieve convergence of the HF energy to within one microhartree and the resulting bond length is $\Req = 2.763$, which is slightly longer than the values in \ce{H2+}, \ce{H2} and \ce{H3+}.
The corresponding energy is $-0.734 337$ which yields a binding energy of $0.2343$ per bond.
In comparison, the binding energy in \ce{H2} is roughly 80\% of this value. This explains the particular stability of the equally-spaced \ce{H_$\infty$} chain in 1D.


\section{Concluding Remarks}
We have studied the electronic structure of 1D chemical systems in which all nuclei and electrons are constrained to remain on a line. 
We have used the full Coulomb operator and our numerical results are strikingly different from those of previous studies\cite{Wagner12, Stoudenmire12} in which a softened operator was used.  We have explored atoms with up to 10 electrons, one- and two-electron diatomics, the chemistry of \ce{H3+} and an infinite chain of \ce{H} atoms.

We find that, whereas atoms with odd numbers of electrons have non-vanishing dipole moments and are reactive, atoms with even numbers of electrons have zero dipole moments and are inert.  Based on these results, we have concluded that the 1D version of the periodic table has only two groups: alkali metals and noble gases.

Our study of one- and two-electron diatomics has revealed that atoms in 1D are bound together by strong one-electron bonds which arise primarily from electrostatic (chiefly charge-dipole and dipole-dipole) interactions.  This leads to a variety of unexpected results, such as the discovery that the bond in \ce{H2+} is much stronger than the bond in \ce{H2}.

\begin{acknowledgments}
P.F.L.~and P.M.W.G.~thank the NCI National Facility for generous grants of supercomputer time.  
P.M.W.G.~thanks the Australian Research Council for funding (Grants No.~DP120104740 and DP140104071). 
P.F.L. thanks the Australian Research Council for a Discovery Early Career Researcher Award (Grant No.~DE130101441) and a Discovery Project grant (DP140104071).
C.J.B. is grateful for an Australian Postgraduate Award.  
\end{acknowledgments}

\appendix

\section{
\label{app:exact}
Some exact wave functions for \ce{H2+}}
The Schr{\"o}dinger equation of a one-electron homonuclear diatomic molecule \ce{A$_2^{Z-1}$} of nuclear charge $Z$ in its state \ce{A1A$^{Z-1}$} is
\begin{equation}
\label{H-H2+}
	-\frac{1}{2} \frac{d^2 \psi(x)}{dx^2} - \left( \frac{Z}{R/2+x} + \frac{Z}{R/2-x} \right) \psi(x) = E\,\psi(x).
\end{equation}
The equation can be solved for $E = 0$, yielding
\begin{equation}
	\psi_n(x) = (1 - z^2) 
	\begin{cases}
		x\,F\left(-\frac{n-1}{2},\frac{n+4}{2},2,1-z^2 \right),
		&
		\text{$n$ odd},
		\\
		F\left(-\frac{n}{2},\frac{n+3}{2},2,1-z^2\right),
		&
		\text{$n$ even},
	\end{cases}
\end{equation}
where $F(a,b,c,x)$ is the Gauss hypergeometric function,\cite{NISTbook} $z=2x/R$ and 
\begin{equation}
	R = \frac{(n+1)(n+2)}{2Z}.
\end{equation}
For example, \ce{H2+} with bond length $R = 1$ has the wave function $\psi_0(x) = (1-2x)(1+2x)$.

\end{document}